\documentstyle[aps,preprint]{revtex}
\input psfig
\begin{document}
\draft
\preprint{UFIFT-HEP-95-6, Fermilab-Pub-95/080-A}
%\preprint{Fermilab-Pub-95/080-A}
\date{April 13, 1995}
\title{The velocity peaks in the cold\\ dark matter spectrum on
Earth}

\author{P. Sikivie$^{1,2}$, I. I.\ Tkachev$^{3,4}$ and Yun
Wang$^{3}$}
\address{
$^{1}$Physics Department, University of Florida, Gainesville, FL
32611}
\address{
$^{2}$Institute for Theoretical Physics, Santa Barbara, CA 93106}
\address{
$^{3}$NASA/Fermilab Astrophysics Center, FNAL, Batavia, IL~~60510}
\address{
$^{4}$Institute for Nuclear Research,
Russian Academy of Sciences, Moscow 117312, Russia}

\maketitle

\begin{abstract}
The cold dark matter spectrum on earth is expected to have peaks in
velocity space. We obtain estimates for the sizes and locations  of
these peaks. To this end we have generalized the secondary infall
model of galactic halo formation to include angular momentum of the
dark matter particles. This new model is still spherically symmetric
and it has self-similar solutions. Our results are relevant to direct
dark matter search experiments.

\end{abstract}


\narrowtext

There is considerable evidence that galaxies are surrounded by dark
halos which contribute 90\% or more of the galactic mass \cite{dmr}.
Identifying the nature of the dark matter is the goal of considerable
theoretical and experimental effort. The leading dark matter
candidates are baryons, neutrinos, weakly interacting massive
particles (WIMPs) and axions.  From the point of view of galaxy
formation, axions and WIMPs have
identical properties which earn them the name of cold dark matter
(CDM).  Unlike baryonic dark matter, CDM is garanteed to have
negligible interactions other than through its gravitational effects.
Unlike neutrinos, CDM has negligibly small primordial
velocity dispersion.  Studies of large scale structure
formation support the view that the dominant fraction of dark matter
is CDM.  Moreover, if some fraction of the dark matter is CDM, it
necessarily contributes to galactic
halos by falling into the gravitational wells of galaxies. The main
focus of our paper is to study the growth of galactic halos by the
infall of cold dark matter, including the effect of angular momentum,
and thence to derive properties of the spectrum of dark matter
particles on Earth.

We are motivated in part by the prospect that a direct search
experiment may some day measure that spectrum.
In particular, if a signal is found in the cavity detector of
galactic halo axions \cite{axd,kvb}, it will be possible to measure
the spectrum with great precision and resolution. Naturally, the
question arises what can be
learned about our galaxy from analyzing such a signal.  Moreover,
knowledge of the spectrum, which is identical for both CDM
candidates, may
help in the discovery of a signal.

In many past discussions of dark matter detection on earth, it has
been assumed that the dark matter particles have an isothermal
distribution. Thermalization has been argued to be the result of a
period of ``violent relaxation'' \cite{lb} following the collapse of
the
protogalaxy. If it is strictly true that the velocity distribution of
the dark matter particles is isothermal, which seems to be a very
strong assumption, then the only information that can be gained from
its observation is the corresponding virial velocity and our own
velocity with respect to its standard of rest. If, on the
other hand, the thermalization is incomplete, a signal in a dark
matter detector may yield additional information.

J. R. Ipser and one of us discussed \cite{is92} the extent to which
cold dark particles are thermalized in a galactic halo and concluded
that there are substantial deviations from a thermal distribution in
that the highest energy particles have narrow peaks in velocity
space. There is one velocity peak on earth due to particles falling
onto the galaxy for the first time, one peak due to particles
falling out of the galaxy for the first time, one peak due to
particles
falling into the galaxy for the second time, and so on. A simple
topological argument shows that these peaks exist no matter what is
the distribution of angular momenta the dark matter particles have
with respect to the galactic center, although angular momentum may
reduce the sizes of the peaks. The peaks due to particles which have
fallen in and out of the galaxy a large number of times in the past
are washed out by scattering in the gravitational
wells of stars, globular clusters and large molecular clouds. Those
particles are in effect thermalized. But the peaks due to particles
which have fallen in and out of the galaxy only a few times in the
past are not washed out by scattering.

The width of the peaks due to
the primordial velocity dispersion of the CDM particles is very
small: $\Delta E/m \sim 10^{-18}$ for axions and $\Delta E/m \sim
10^{-14}$ for WIMPs, where $m$ is the mass of the particles and
$\Delta E$ their energy spread.  As just mentioned, the peaks
fragment
because of the gravitational scattering of the particles by
inhomogeneities in the galaxy. They also fragment because of
structure formation on scales smaller than that of the galaxy as a
whole. We expect that such structure formation in the CDM component
of the matter density would be inhibited by the tidal forces of the
galaxy's gravitational field, but some of it no doubt occurs. At any
rate, it is clear that the peaks due to particles which have fallen
in and out of the galaxy only a small number of times in the past are
not entirely washed out by either scattering or small structure
formation. It is the purpose of the present  paper to estimate the
sizes and locations of these peaks and we will no longer concern
ourselves with their widths.  Let us point
out however that the
sensitivity of the search for galactic halo axions using the cavity
detector may be increased by looking for narrow peaks. Specifically,
it is found that the signal to noise ratio of the upcoming axion
search
at LLNL \cite{kvb} is increased by a factor $180 f_1$ by looking for
narrow
peaks, where $f_1$ is the fraction of the local axion density in the
largest of the peaks of width $\Delta E < 10^{-11}\, m$.

The tool we use to obtain estimates of the
locations and sizes of the highest energy peaks is the secondary
infall model of galactic halo formation \cite{gunn}. In its original
form, this model makes the following assumptions:

1) the dark matter is non-dissipative; 2) it has
negligible initial velocity dispersion; 3) the gravitational
potential of the galaxy is spherically symmetric and is dominated by
the dark matter contribution; 4) the dark matter particles have zero
angular momentum and therefore move on radial orbits through the
galactic center.

Below, we will generalize the model to rid it of the fourth
assumption.
For clarity, we refer to the model with the fourth assumption
included
as the radial infall model.

An initial overdensity profile $\delta M_i (r)$ is assumed. The
equations of motion for the radial coordinate $r(\alpha, t)$ of each
spherical shell ($\alpha$ is a shell label, t is time) in the
gravitational potential due to all the other shells must then be
solved for initial conditions given by the Hubble expansion at some
arbitrarily chosen but early time $t_i$: $\dot{r}(\alpha, t_i)=H(t_i)
r(\alpha, t_i)$.  Much progress in the analysis of the model
came about as a result of the realization that
the evolution of the galactic halo is self-similar \cite{fg} provided
the initial overdensity has the following scale-free form:
\begin{equation}
{\delta M_i \over M_i} = \left({M_0 \over M_i} \right)^\epsilon
\,\, ,
\label{inm}
\end{equation}
where $M_i$ and $\delta M_i$ are respectively the mass and excess
mass interior to $r_i$ at the initial time $t_i$, i.e.
\begin{equation}
M_i(r_i)={4\pi \over 3} \rho(t_i)r_i^3 +\delta M_i(r_i) \, ,
\label{mi2}
\end{equation}
where $\rho(t_i)= {3 H(t_i)^2 / 8\pi G} ={1 /6\pi Gt_i^2}$ is the
average density in a critical $(\Omega =1)$ universe,
$t_i$ having been chosen during the matter-dominated epoch. $M_0$ and
$\epsilon$ are constants characterizing the model. Self-similarity
means that the phase-distribution of dark matter particles is
time-independent after all distances have been rescaled by the
overall size $R(t)$ of the galactic halo, all masses by the mass
$M(t)$ interior to the radius $R(t)$, and all velocities by
$\sqrt{GM(t)/R(t)}$. For the sake of definiteness, $R(t)$ is taken to
be the ``turn-around'' radius at time $t$, i.e. the radius at which
particles have zero radial velocity for the first time in their
history (see Fig.1). In a self-similar solution, the motion of each
shell $\alpha$ is the same after the appropriate rescaling, i.e.
$r(\alpha, t)=r_*(\alpha)\lambda(t/t_*(\alpha))$, where $r_*(\alpha)$
and $t_*(\alpha)$ are respectively the turn-around radius and time of
shell $\alpha$. Also the mass-profile of the halo is time-independent
after rescaling, i.e. $M(r,t)=M(t){\cal M} (r/R(t))$. $\lambda
(\tau)$ and ${\cal M} (\xi)$ are functions of a single variable which
can be accurately obtained by numerical integration. In the limit
$\xi \rightarrow 0$ it was shown analytically \cite{fg} that, ${\cal
M}(\xi) \propto \xi$ if $0 \le \epsilon \le 2/3$
and ${\cal M}(\xi) \propto
\xi^{3/(1+3\epsilon)}$ if $2/3 \le \epsilon \le 1$. The numerical
integrations confirm this. Thus, for the range $0 \le \epsilon \le
2/3$, the radial infall model produces flat rotational curves, i.e.
it is in accord  with the main feature of the galactic mass
distribution.

To fit the model to our present galactic halo we must
determine appropriate values of $R(t_0)$ and $M(t_0)$, where $t_0$ is
the age of the universe. First, we have:

\begin{equation}
t_0={\pi \over 2} \sqrt{R(t_0)^3 \over 2GM(t_0)}={6.52 \times 10^9
\text{year} \over h} \, ,
\label{t0}
\end{equation}
where $h$ parametrizes the present Hubble rate: $H_0 =h \, 100 $ km
sec$^{-1}$ Mpc$^{-1}$. The first equality in Eq.(\ref{t0}) follows
from the fact that a given shell does not cross any other shell till
after the moment of its first turnaround and therefore the mass
interior to it stays constant till then. The second equality in Eq.
(\ref{t0}) follows from the assumption $\Omega =1$ which is necessary
to have a self-similar solution. Second, we match the rotation
velocity in the model to the observed one for our galaxy:

\begin{equation}
v_{\rm rot}=\nu (\epsilon) \sqrt{GM(t_0) \over R(t_0)}= 220\,
\text{km}\, \text{sec}^{-1}  \, ,
\label{vr}
\end{equation}
where $\nu (\epsilon)$ is extracted from the numerical solution of
the model. Combining Eqs. (\ref{t0}) and (\ref{vr}), we have:
\begin{equation}
R(t_0)={1.32\, \text{Mpc} \over h\nu (\epsilon)}   \, .
\label{Rt}
\end{equation}
For typical values of $\epsilon$ and $h$, $R(t_0)$ turns out to be in
the 1 to 3 Mpc range; see Table I. This is consistent with the value
one
would infer by taking nearby galaxies as tracers of
mass. Indeed M31, at a distance of 0.73 Mpc, is falling towards us
with a line-of-sight velocity of 120 km/s, whereas galaxies at
distances exceeding 3 Mpc are receding from us as part of the
universal Hubble expansion. In fact, turning this around, Eqs.
(\ref{vr}) and (\ref{Rt}) could form a basis for estimating the age
of the universe.

Figs.1 and 2 show the phase-space diagram and the velocity peaks on
earth for $\epsilon =0.2$ and $h=0.7$. Our distance to the galactic
center is taken to be $r_{\rm S}=8.5$ kpc. The rows labeled $j=0.0$
in
Table I give the density fractions and kinetic energies of the first
five incoming peaks in the radial infall model. For each
incoming peak there is an outgoing peak with approximately the same
energy and density fraction (see Fig.2). Note that, because of the
scale
invariance of the model, increasing
$h$ at fixed $\epsilon$ is equivalent to decreasing $R(t_0)$
keeping the observation radius $r_{\rm
S}$ fixed, {\rm or} to increasing $r_{\rm S}$ at fixed $R(t_0)$.
We find that, in the radial infall model, the sizes of the
two peaks due to particles falling in and out of the galaxy for the
first time are large, each containing of order 10\% of the
local halo density for $\epsilon$ in the standard CDM model inspired
\cite{hs} range of 0.1 - 0.3.

However, because the radial infall model neglects the angular
momentum that the dark matter particles are expected to have, we
must question its reliability in this context. In particular, since
in that model all particles go through the galactic center at each
pass,
the halo density is large there, behaving as
$\rho_{\rm halo} (r) \sim 1/r^2\sqrt{\ln (1/r)}$. Instead, the actual
halo mass distribution has the approximate form $\rho_{\rm halo} (r)
= \rho_{\rm halo} (0)/(1+(r/a)^2)$, where $a$, called the core
radius, is of order a few kpc \cite{bt}. Since $r_{\rm S}$ and $a$
have the
same order of magnitude, the corrections due to angular momentum
are not small.  Moreover, angular momentum affects
most the peaks with the highest energies (the ones we are most
interested in) because these are the ones due to particles which have
come from furthest away.

Fortunately, there is a generalization of the radial infall model
which takes angular momentum into account while still keeping the
model tractable. In this generalization, each shell $\alpha$ is
divided into $N$ subshells labelled by an index $k$, (k=1, ... , N).
The particles in a given subshell $(\alpha, k)$ all have the same
{\it magnitude} $l_k (\alpha)$ of angular momentum. At each point on
each subshell, the distribution of angular momentum vectors is
isotropic about the axis from that point to the galactic center. Thus
the spherical symmetry of each subshell is maintained in time.
Moreover, the evolution is self-similar provided:

\begin{equation}
l_k (\alpha)=j_k r_*(\alpha)^2/t_*(\alpha)   \, ,
\label{lk}
\end{equation}
where the $j_k$ are a set of dimensionless numbers characterizing the
galaxy's angular momentum distribution. In Eq.(\ref{lk}) we are
neglecting the small dependence  of the turn-around radius
$r_*(\alpha)$ and time $t_*(\alpha)$ upon $k$, although that neglect
is not necessary for self-similarity. The model will be described in
detail elsewhere \cite{stw}.  Let us just give here the two
equations:
%\FL
\begin{mathletters}
\label{ml}
\begin{equation}
{d^2\lambda_k \over d\tau^2}= {j_k^2 \over \lambda_k^3}-{\pi^2
\over 8}{\tau^{2/3\epsilon} \over \lambda_k^2} {\cal M}\left(
{\lambda_k \over \tau^{2/3 +2/9\epsilon} }\right) \,\, ,
\label{mla}
\end{equation}
%\FL
\begin{equation}
{\cal M}(\xi ) = {2\over 3\epsilon }\sum _{k=1}^N n_k
\int_1^{\infty}{d\tau
\over \tau^{1+2/3\epsilon}} \theta\left(\xi -{\lambda_k (\tau ) \over
\tau^{2/3+2/9\epsilon}}\right) ,
\label{mlb}
\end{equation}
\end{mathletters}
and the boundary conditions: $\lambda_k(1)=1$,
$d\lambda_k/d\tau(1)=0$, which determine the model's evolution
through $r_k(\alpha, t)=r_*(\alpha)\lambda_k(t/t_*(\alpha))$ and
$M(r,t)=M(t){\cal M} (r/R(t))$. In Eq. (\ref{mla}), $n_k$ is the mass
fraction
contributed by the subshell $k$. In all cases presented here, the
$j_k$ are taken to be distributed according to the density
\begin{equation}
{dn \over dj}={2j \over j_0^2} \exp(-j^2/j_0^2)  \,\, .
\label{nj}
\end{equation}
We report our results in terms of the average
$\bar{j}=\sqrt{\pi}j_0/2$. Fig.3 shows the rotation curves for the
case
$\epsilon=0.2$,
$j=0$ (whose phase-space diagram and velocity peaks are shown in
Figs.1 and 2) and the case $\epsilon=0.2$, $\bar{j}=0.2$. It shows
that
the effect of angular momentum is to give a core radius to the
galactic halo distribution.  The definition of core radius we use
below to report our results
is the radius $b$ at which half of the rotation velocity squared is
due
to the halo. For the density profile $\rho(r)=\rho(0)/(1+(r/a)^2)$,
$b=2.33 a$; of course, our density profiles are only
qualitatively similar to that one.  Fig. 4 shows the velocity peaks
for the case $\epsilon=0.2$, $\bar{j}=0.2$, $h=0.7$. Table I gives
the values of the current turn-around radius $R(t_0)$, the core
radius
$b$, the halo density at our location $\rho(r_{\rm S})$,  and the
density fractions and energies of the five most energetic incoming
peaks for various values of $\epsilon$, $\bar{j}$ and $h$.
The range of $\epsilon$
values chosen is motivated by models of large-scale structure
formation \cite{hs}  as well as by the flatness of the rotation
curves produced. Table I is the summary of our results.

In conclusion, we have generalized the radial infall model of
galactic halo formation to include the effect of angular momentum,
all the while keeping the model spherically symetric and self-similar
in its time evolution.  The galactic halo distributions we obtain
have core radii as well as flat rotation curves. We find that the
contribution to the local halo density due to particles which are
falling in and out of the galaxy for the first time or which have
passed through the galaxy only a small number of times in the past,
and which are therefore not thermalized, is rather large, comprising
several percent per velocity peak. Finally let us emphasize
that the peak sizes we
obtain are only order of magnitude estimates since they are averages
over a broad distribution of possible angular momenta.

\acknowledgments
We thank S. Colombi, J. Ipser, J. Primack and D. Spergel for useful
discussions.
This research was supported in part by the DOE and NASA grant
NAGW-2381
at Fermilab, and the DOE grant DE-FG05-86ER40272 at the University of
Florida.

%%%%%%%%%%%%%%%%%%%%%%
%\newpage
\widetext

\begin{table}
\caption{Relative magnitudes $A_k$ and kinetic energies $E_k$ of the
first five incoming peaks for various values of $\epsilon$, $\bar{j}$
and $h$.
Also shown are the current turn-around radius $R$ in
units of Mpc, the core radius $b$ in kpc, and the local density
$\rho$ in
$10^{-25}$ g cm$^{-3}$. The $A_k$ are in percent and the
$E_k$ are in units of $0.5 \times (300 $ km s$^{-1})^2$.}
\begin{tabular}{ccccccccccc}
 $\epsilon$ &  $\bar{j}$ & $h$ & $R$ & $b$ & $\rho$  &$A_1\,\, (E_1)$
&
$A_2\,\, (E_2)$ &
$A_3\,\, (E_3)$ & $A_4\,\, (E_4)$ & $A_5\,\, (E_5)$   \\
\tableline
 0.2  & 0.0  &  0.7 & 2.0  & 0.0 & 8.1 & 13 ~(4.0) & 5.3 ~(3.2) & 3.3
{}~(2.7) & 2.4 ~(2.4) & 1.9 ~(2.2)\\
 1.0  & 0.0  &  0.7 & 0.9  & 0.0  & 8.4 & 1.6 ~(3.4) & 1.1 ~(3.2) &
0.9
{}~(3.0) &  0.8 ~(2.9) &  0.7 ~(2.8)\\
\tableline
 0.15  & 0.2   &  0.7 & 2.4   & 13  & 5.0  & 4.0 ~(3.1) & 5.4 ~(2.3)
& 5.3
{}~(1.8) & 4.9 ~(1.5) & 4.0 ~(1.3)\\
 0.2   & 0.1   &  0.7 & 2.0   & 4.5  & 7.6 & 7.4 ~(3.8) & 7.2 ~(3.0)
& 4.9
{}~(2.5) & 3.2 ~(2.2) & 2.4 ~(2.0)\\
 "   & 0.2   &  0.7 & 2.0  & 12  & 5.4 & 3.1 ~(3.4) & 4.1 ~(2.6) &
4.3 ~(2.1) &
4.1 ~(1.8) & 3.6 ~(1.6)\\
 "   & "   &  0.5 & 2.8 & 17 & 4.9 & 1.9 ~(3.5) & 2.5 ~(2.7) & 2.8
{}~(2.3) & 2.9
{}~(2.0) & 3.0 ~(1.7)\\
 "   & "   &  0.9 & 1.6  & 9.3 & 6.0 & 4.4 ~(3.2) & 5.3 ~(2.5) & 5.1
{}~(2.0) &
4.5 ~(1.7) & 3.6 ~(1.5)\\
 "   & 0.4   &  0.7 & 2.0   & 40  & 2.6 &  0.8 ~(2.5) & 1.6 ~(1.8) &
2.1 ~(1.4)
& 2.4 ~(1.1) & 2.6 ~( 0.9)\\
 0.25  & 0.2  &  0.7 & 1.8 & 8.5 & 5.5 & 2.0 ~(3.5) & 2.9 ~(2.8) &
3.3 ~(2.4) &
3.4 ~(2.1) & 3.1 ~(1.8)\\
 0.4  & 0.2  &  0.7 & 1.5 & 2.2 & 7.7 & 1.1 ~(4.0) & 1.5 ~(3.4) & 1.8
{}~(3.0) &
1.9 ~(2.8) & 2.1 ~(2.5)\\
\end{tabular}
\label{tbl2}
\end{table}

\narrowtext

\begin{figure}
\psfig{file=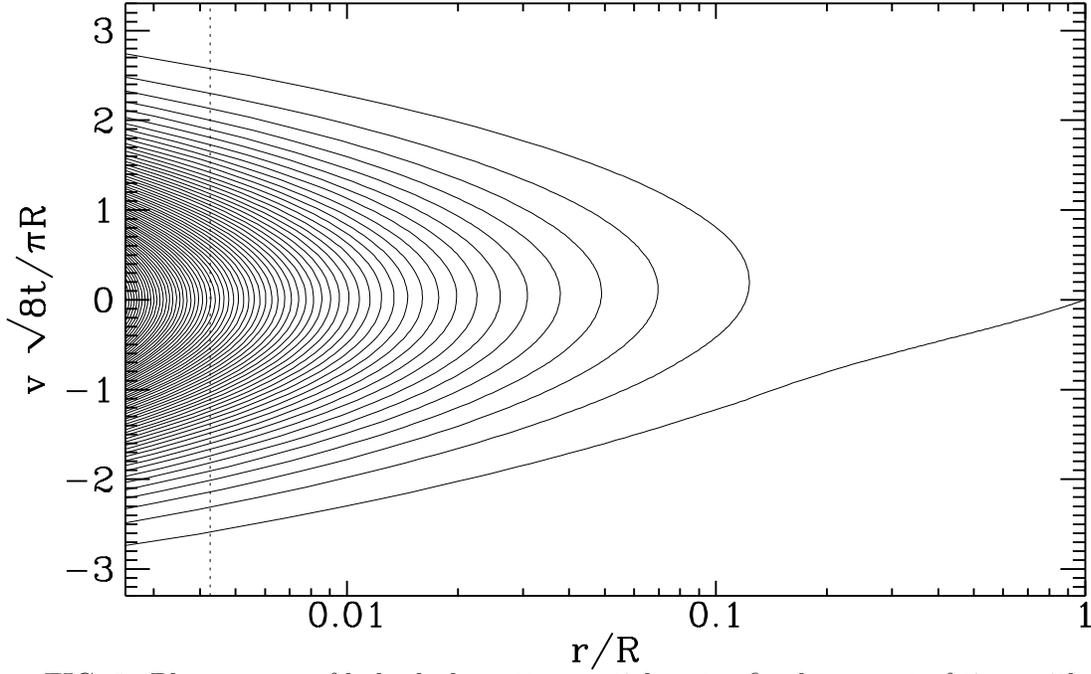,height=3.8in,width=6in}
\caption{Phase space of halo dark matter particles at a fixed moment
of
time with
$\epsilon =0.2$ and $j=0$. Dotted line corresponds to the Sun's
position
if $h=0.7$.}
\label{fig:phs}
\end{figure}

\begin{figure}
\psfig{file=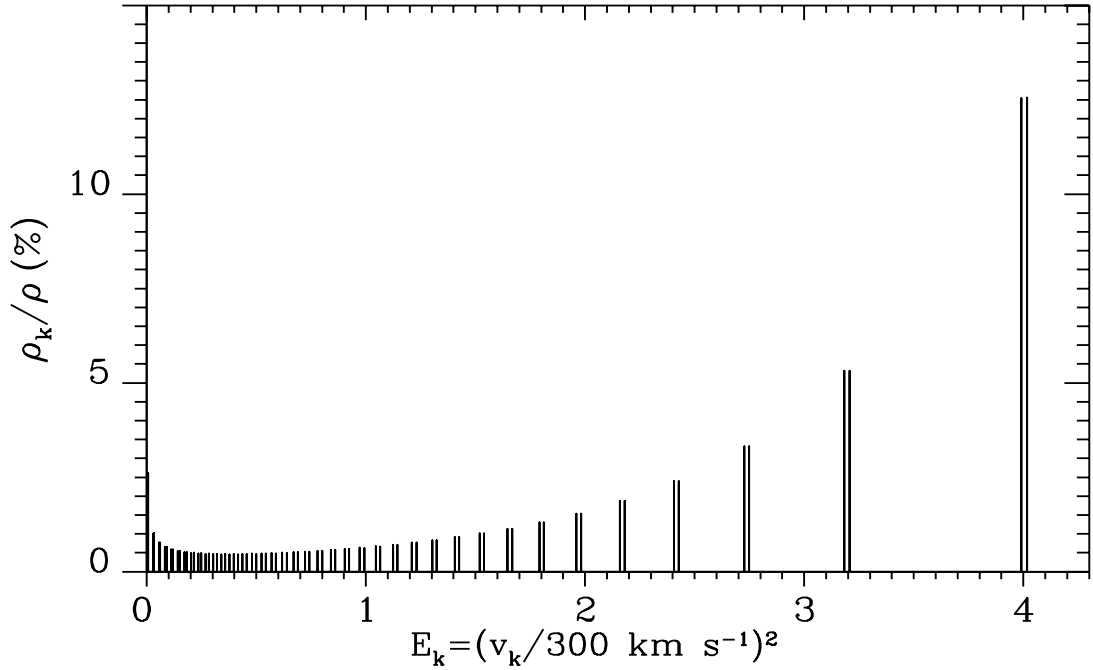,height=3.7in,width=6in}
\caption{The spectrum of velocity peaks at the Sun's position for the
case
$\epsilon
=0.2$, $j=0$ and $h=0.7$.}
\label{fig:frs}
\end{figure}

\begin{figure}
\psfig{file=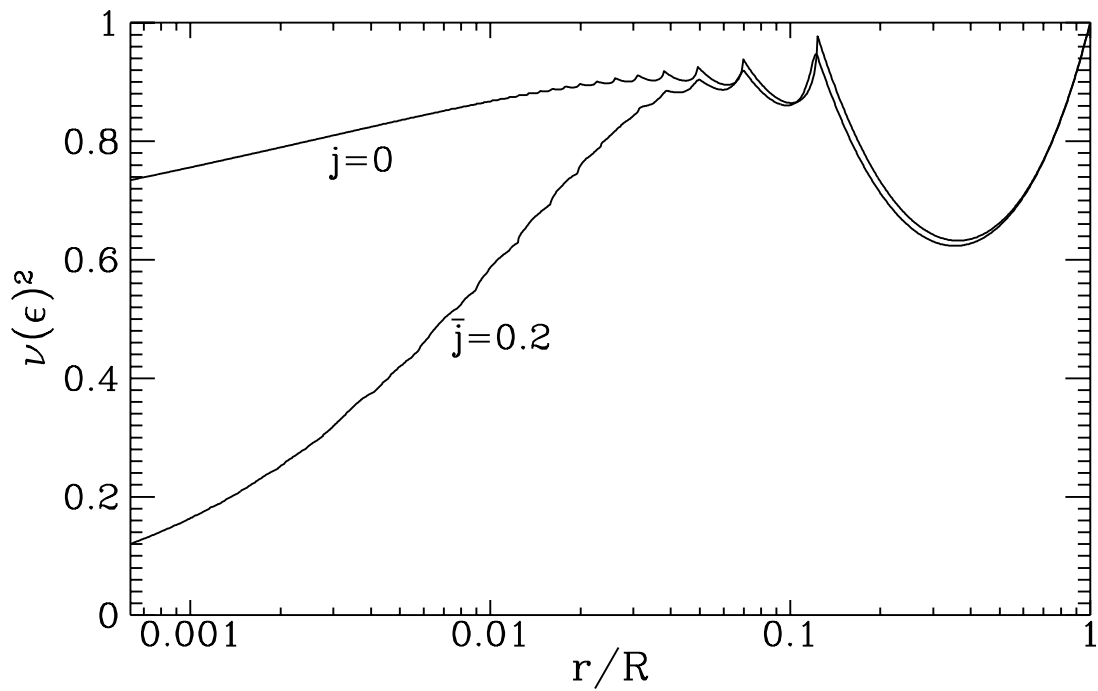,height=3.8in,width=6in}
\caption{Rotational curves for the case $\epsilon =0.2$, with and
without angular momentum.}
\label{fig:rvj02}
\end{figure}

\begin{figure}
\psfig{file=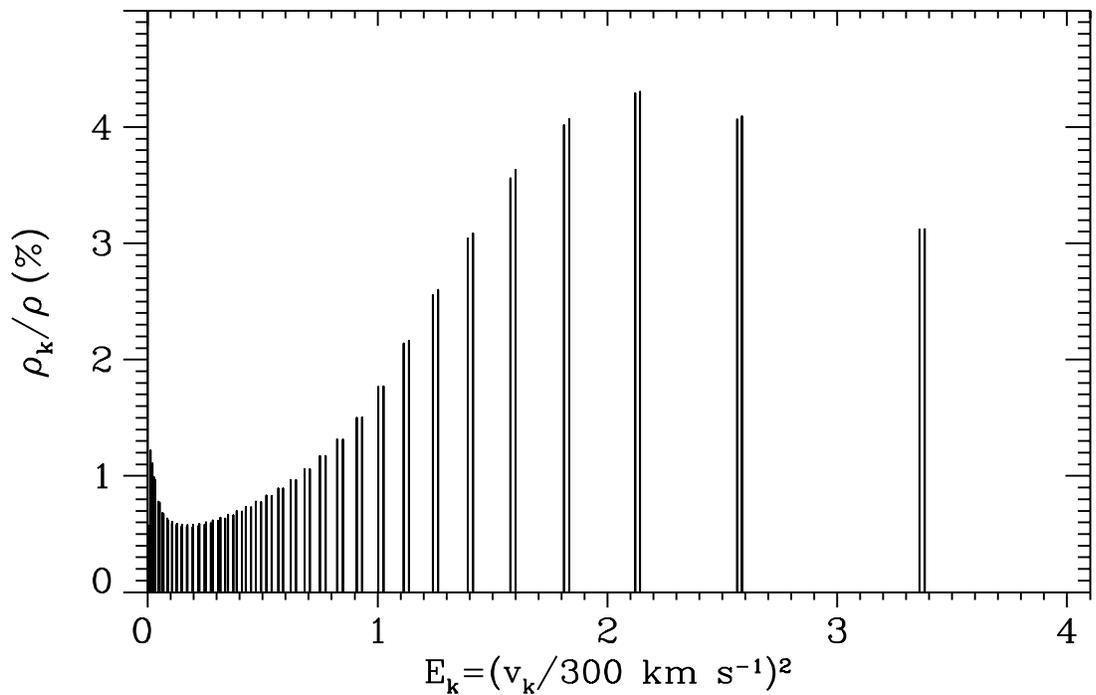,height=3.8in,width=6in}
\caption{The spectrum of velocity peaks for the case $\epsilon =0.2$,
$\bar{j}=0.2$ and $h=0.7$.}
\label{fig:vp}
\end{figure}

\end{document}